\documentclass[12pt]{article}
\usepackage[bottom=0.8in]{geometry}

\usepackage{authblk} 
\usepackage{amsthm,amsmath,amssymb} 
\usepackage{amsmath} 
\usepackage{amsfonts}
\usepackage{subcaption} 
\usepackage{float} 
\usepackage{color}

\usepackage{enumerate}
\usepackage{natbib} 
\usepackage{url} 
\usepackage{setspace} 
\usepackage{graphicx,psfrag,epsf} 

\begin{document}

\newcommand{\PM}{PM$_{2.5}$}
\newcommand{\mugm}{$\mu$g/m$^3$}
\newcommand{\indep}{\rotatebox[origin=c]{90}{$\models$}}

\def\spacingset#1{\renewcommand{\baselinestretch}%
{#1}\small\normalsize} \spacingset{1}

\newtheorem{assumption}{Assumption}
\setcounter{assumption}{4}

\title{Causal exposure-response curve estimation with surrogate confounders: a study of air pollution and children's health in Medicaid claims data}

\author[1]{Jenny J. Lee} 
\author[2]{Xiao Wu}
\author[1]{Francesca Dominici} 
\author[1]{Rachel C. Nethery \thanks{rnethery@hsph.harvard.edu}}
\affil[1]{Department of Biostatistics, Harvard T.H. Chan School of Public Health, Boston, MA }
\affil[2]{Department of Biostatistics, Mailman School of Public Health, Columbia University, New York, NY}
\date{} 
\maketitle

\begin{abstract}
{In this paper, we undertake a case study to estimate a causal exposure-response function (ERF) for long-term exposure to fine particulate matter (PM$_{2.5}$) and respiratory hospitalizations in socioeconomically disadvantaged children using nationwide Medicaid claims data. These data present specific challenges. First, family income-based Medicaid eligibility criteria for children differ by state, creating socioeconomically distinct populations and leading to clustered data. Second, Medicaid enrollees' socioeconomic status,  a confounder and an effect modifier of the exposure-response relationships under study, is not measured. However, two surrogates are available: median household income of each enrollee's zip code and state-level Medicaid family income eligibility thresholds for children. We introduce a customized approach for causal ERF estimation called \textit{MedMatch}, building on generalized propensity score (GPS) matching methods. \textit{MedMatch} adapts these methods to (1) leverage the surrogate variables to account for potential confounding and/or effect modification by socioeconomic status and (2) address practical challenges presented by differing exposure distributions across clusters. We also propose a new hyperparameter selection criterion for \textit{MedMatch} and traditional GPS matching methods. Through extensive simulation studies, we demonstrate the strong performance of \textit{MedMatch} relative to conventional approaches in this setting. We apply \textit{MedMatch} to estimate the causal ERF between PM$_{2.5}$ and respiratory hospitalization among children in Medicaid, 2000-2012. We find a positive association, with a steeper curve at lower PM$_{2.5}$ concentrations that levels off at higher concentrations.}
\end{abstract}


\newpage
\spacingset{1.45} 

\section{Introduction}\label{Section:Introduction}

In the United States, the Clean Air Act explicitly requires the National Ambient Air Quality Standards to be sufficient to protect the health of sensitive populations, such as socioeconomically disadvantaged children. Children are known to be more susceptible to air pollution exposure as it can disrupt their respiratory, neurological, and immune systems, which are still developing and immature. It is relatively well known that exposure to ambient air pollution, and in particular to fine particulate matter (\PM), has adverse health effects on children \citep{who2018}. However, there has been minimal characterization of the adverse effects of long-term \PM\ exposure on socioeconomically disadvantaged children, who may be more vulnerable to its effects due to increased prevalence of underlying diseases, or inadequate resources to implement precautionary measures \citep{cortes2021environmental}.

\begin{figure}
\centering
 \begin{subfigure}[b]{\textwidth}
	\centerline{
	\includegraphics[width=\textwidth,keepaspectratio]{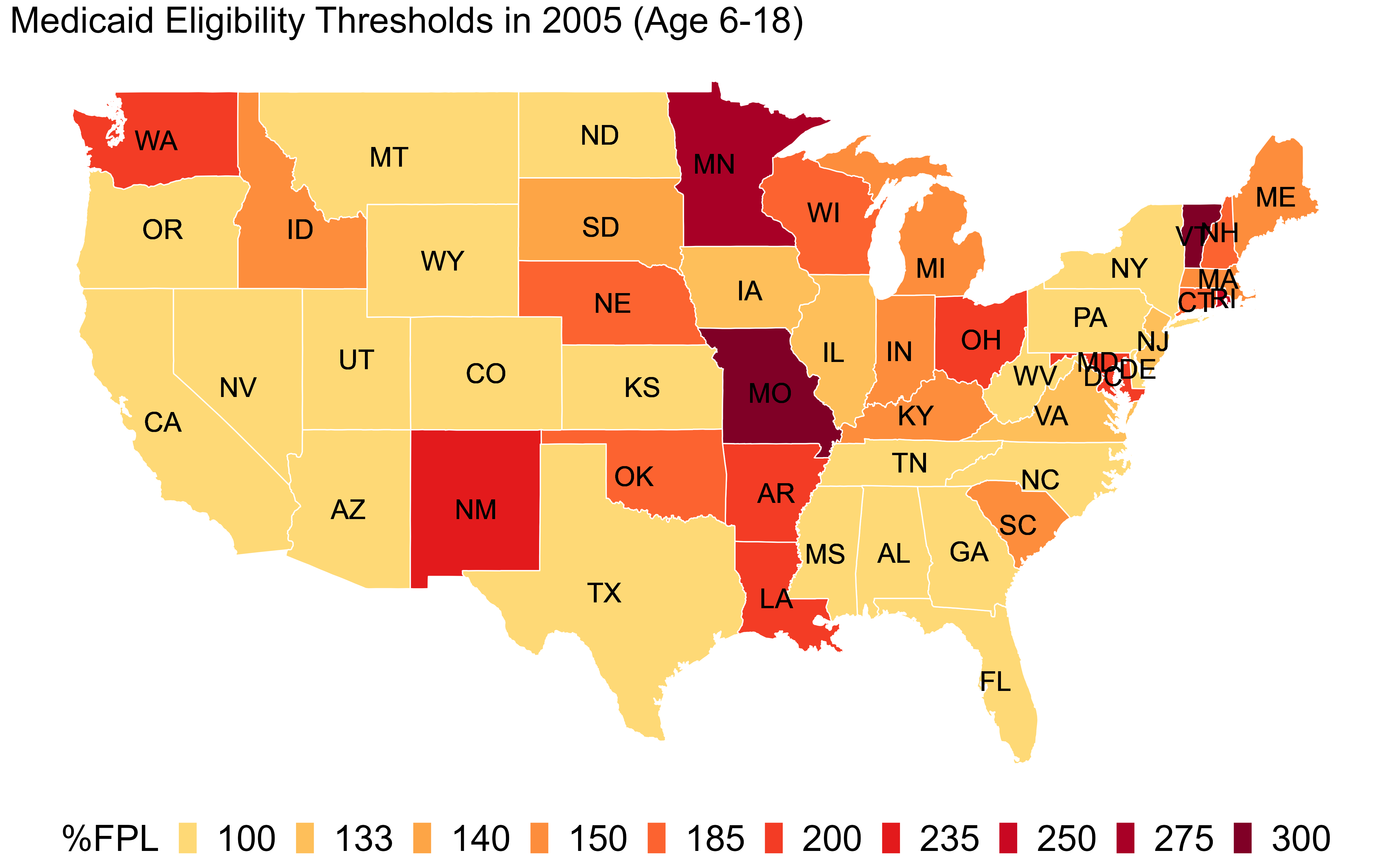}}
	\caption{Map of state Medicaid eligibility thresholds}
 	\label{Figure:US_map_criteria_Age6to18_2005}
  \end{subfigure}
  \hfill
   \begin{subfigure}[b]{\textwidth}
\centerline{
\includegraphics[width=\textwidth,keepaspectratio]{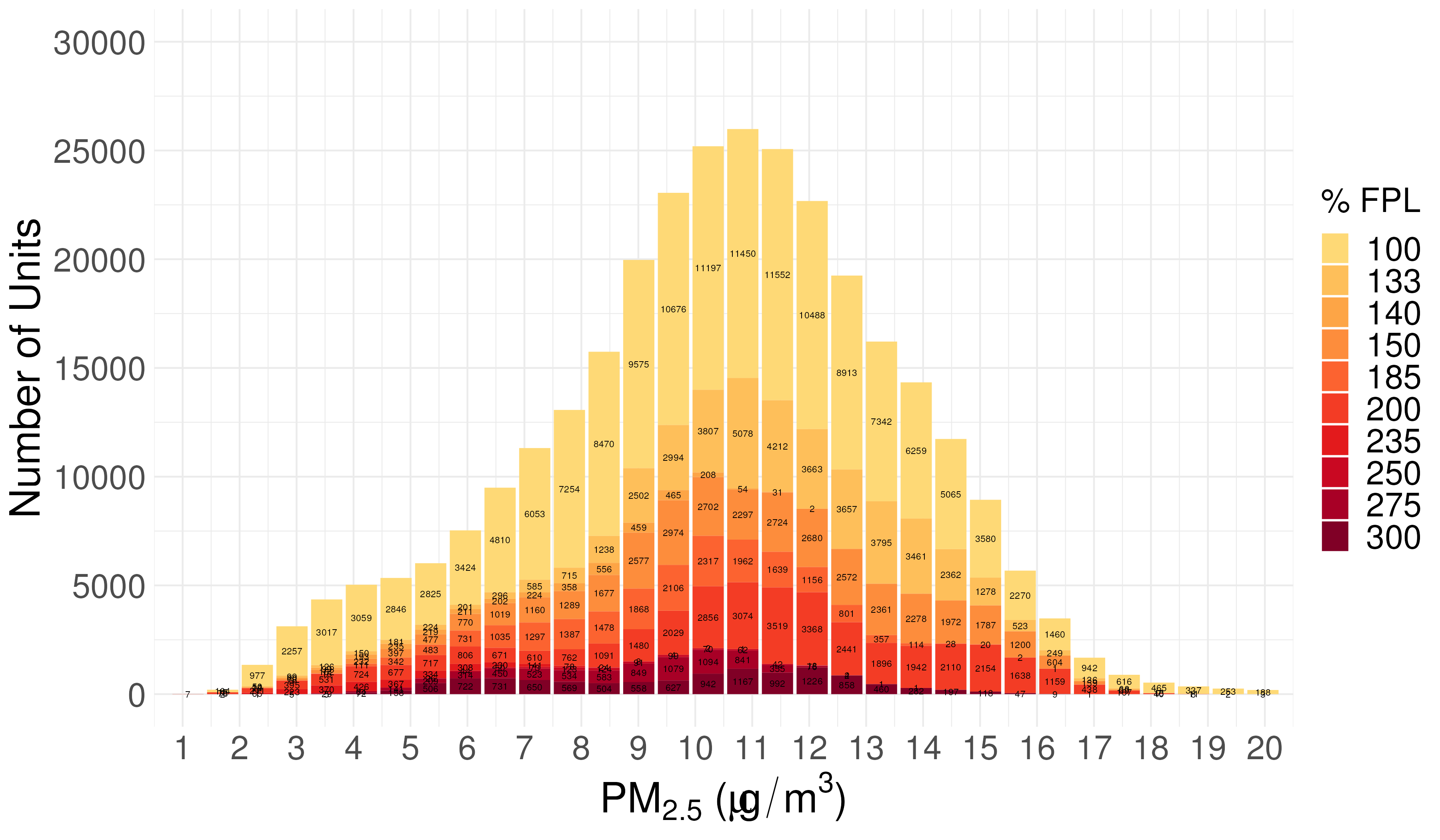}
}
\caption{\PM\ distribution by state Medicaid eligibility threshold}
\label{Figure:StackedHist}
   \end{subfigure}
   \caption{(a) shows a map of each state's Medicaid family income eligibility thresholds for children of age 6-18, calculated as a percentage of the federal poverty level (\%FPL) in 2005. Higher \%FPL indicates a less stringent eligibility threshold. (b) shows a histogram of the distribution of \PM\ exposures across study units (zip codes) during the study period, stratified by state Medicaid eligibility threshold.}
   \label{Figure:motivatingdata}
\end{figure}

In this paper, we study the effect of long-term \PM\ exposure on respiratory hospitalization in low-income children using claims data from the US Medicaid program from 2000-2012. Medicaid insures nearly 40\% of US children, who must come from a low-income family or be disabled to qualify \citep{truffer2016actuarial}. The Medicaid program is jointly funded by the state and federal governments. However, the program is administered by the states, and each state can determine how low a family's income must be in order for the children to be Medicaid eligible (the ``eligibility threshold'') (Figure~\ref{Figure:US_map_criteria_Age6to18_2005}). Therefore states with differing Medicaid eligibility thresholds will have socioeconomically distinct Medicaid populations. As a result, Medicaid claims data can be viewed as having a clustered structure where units (either individuals or small areas) are nested within a state or within a set of states with the same Medicaid eligibility threshold. In our motivating application, zip codes are the units of analysis, and they are nested within clusters defined as sets of states sharing a common Medicaid eligibility threshold. 

Income is known to be associated with both \PM\ and respiratory outcomes, i.e., it is a confounder of the association of interest \citep{greenland1999confounding,hajat2021confounding}. In addition, the health effects of \PM\ vary in magnitude by income levels, i.e., income is an effect modifier \citep{vanderweele2007four,cakmak2016modifying,hajat2021confounding}. Even in Medicaid-enrolled children, a population that is generally low-income, substantial variability in income remains; thus, it is likely to exert a confounding and effect modifying influence. However, Medicaid claims data do not provide individual-level information on the income of the beneficiaries (when using data for children, we would like to know each child's \textit{household income}). However, we have available two useful surrogates: 1) the median household income of the zip code of residence of each Medicaid enrollee (note that this is the median across the entire zip code population, not restricted to Medicaid enrollees in the zip code); and 2) the state-level Medicaid family income eligibility thresholds for children, calculated as a percentage of the federal poverty level \citep{brooks2019medicaid}.

To address these challenges, in this paper we propose a causal inference method for clustered data with surrogate measures called \textit{MedMatch}. \textit{MedMatch} is a customized approach that can estimate a causal exposure-response function (ERF) for \PM\ in Medicaid claims data, accounting for the potential confounding and effect modifying influence of income for Medicaid-enrolled children using the two surrogate measures described above. In addition to allowing for the use of surrogates, \textit{MedMatch} is tailored to combat biases in ERF estimation caused by cluster-patterned data sparsity in some areas of the exposure range.
 
\begin{figure}
	\centerline{
 \includegraphics[width=\textwidth,height=\textheight,keepaspectratio]{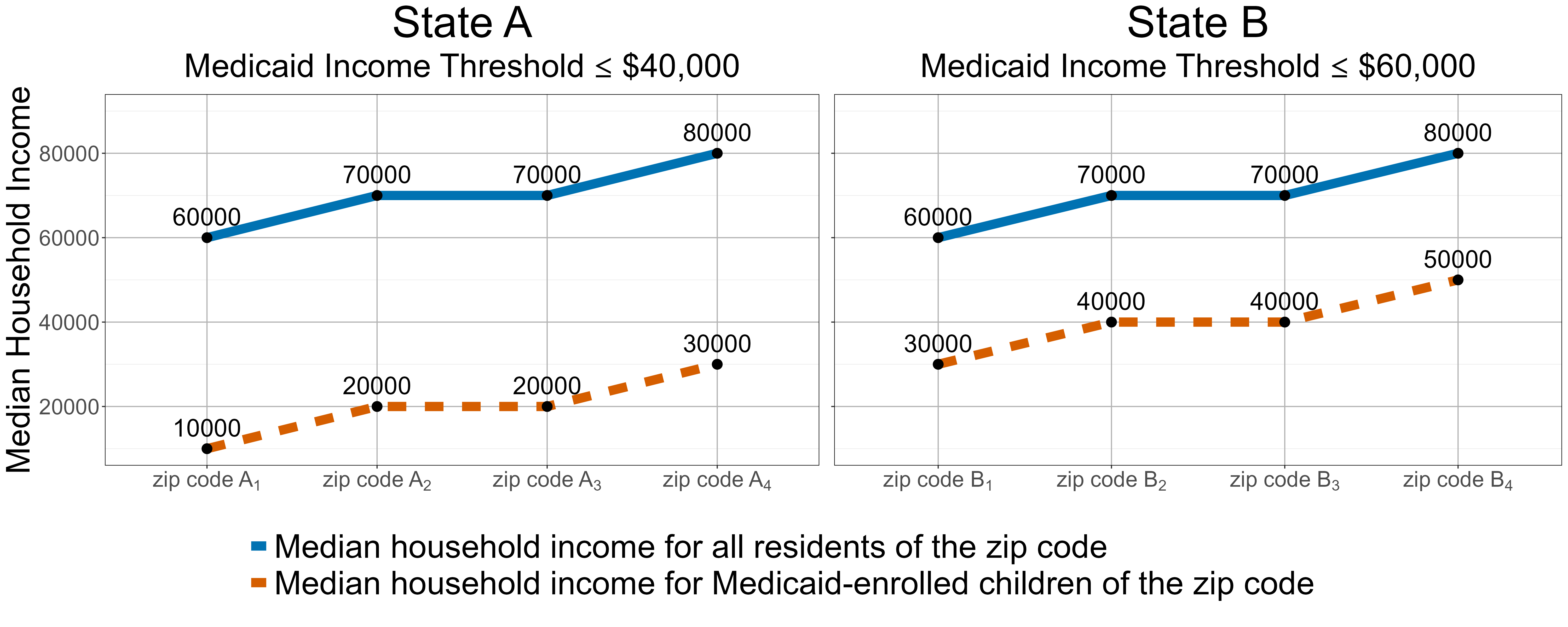}}
	\caption{A toy example of the assumed relationship between median household income for all residents of the zip code (MHI) and median household income of Medicaid-enrolled children in the zip code (MHI-MC), for two states with different Medicaid eligibility thresholds.}
    \label{Figure:Relationship_between_Medicaid_Census_Income}
\end{figure}

We posit two critical assumptions about the structure of the relationship between the unmeasured income variable and the surrogate variables, which will be leveraged by \textit{MedMatch}. To better contextualize these assumptions, as zip codes serve as units of analysis here, we conceive of all confounders/effect modifiers in their zip code-aggregate version in practice. Thus, the ideal income measure (which is unobserved in our data) is the median household income for Medicaid-enrolled children (MHI-MC) within each zip code. Then, our first assumption is that a state's eligibility threshold is an upper bound of the MHI-MC of each of its zip codes. This assumption follows directly from the stated policy that only individuals with income less than their state's threshold are eligible to enroll, thus the MHI-MC must also be less than the threshold. Second, we assume that the zip code median household income for all residents of the zip code (MHI) is a rank-preserving function of the zip code MHI-MC \textit{within states with the same eligibility threshold}. For example, within State A, if zip code $A_1$ has a higher MHI than zip code $A_2$, zip code $A_1$ should also a have higher MHI-MC than zip code $A_2$ (Figure~\ref{Figure:Relationship_between_Medicaid_Census_Income}). However, for two zip codes located in states with different eligibility thresholds, we do not make this rank-preserving assumption (Figure~\ref{Figure:Relationship_between_Medicaid_Census_Income}). For example, consider a zip code $B_1$ from state B and zip code $A_1$ from state A, where state B has higher eligibility threshold than state A. Even if the MHI in $B_1$ and $A_1$ are the same, we expect that the MHI-MC in $B_1$ will be larger because higher-income children in zip code $B_1$ will be Medicaid eligible thanks to the higher eligibility threshold.

Under these assumptions on the relationship between the unmeasured confounder/effect modifier and the surrogates, we show that theoretically a causal ERF can be identified by conditioning on a GPS constructed using the surrogates. 
Only a few recent papers have begun to propose causal inference methods for estimating an ERF. Existing approaches include regression adjustment for the estimated generalized propensity score (GPS) in the outcome model \citep{hirano2004propensity}; a doubly robust kernel smoothing approach \citep{kennedy2017non}; GPS matching \citep{wu2022matching}; and Gaussian processes \citep{ren2021bayesian}. None of these methods explicitly deals with clustered data.

Moreover, while theoretically viable, the ability of these ERF estimation methods to eliminate confounding bias is often compromised in practice due to data sparsity in some regions of the exposure range. This is likely to occur even with very large datasets and when causal identifying assumptions are upheld. We describe how this issue, and likely the accompanying biases, may be exacerbated in clustered data settings. In our own application, due to geographic patterns in \PM\ and Medicaid eligibility, the empirical distribution of \PM\ differs systematically across Medicaid eligibility clusters, with some clusters having few/no observations in certain areas of the exposure range (Figure~\ref{Figure:StackedHist}). This creates a pronounced data sparsity issue, which we refer to as ``cluster-patterned data sparisty'', that can amplify residual confounding biases of existing ERF estimation methods.

To address these issues,
we propose a new ERF estimation method called \textit{MedMatch}, which builds on GPS matching  \citep{wu2022matching}. In particular, it adapts GPS matching to account for the confounding and/or effect modifying influence of unmeasured MHI-MC in our analysis by carefully utilizing the surrogates and their assumed relationships with MHI-MC, while also combating biases due to cluster-patterned data sparsity. \textit{MedMatch} has the following two key features: 1) it uses the MHI-MC surrogates to construct the GPS and 2) it introduces a two-part matching target that seeks matches with comparable GPS values while also promoting matches within Medicaid eligibility clusters or across clusters with similar eligibility thresholds. 
We evaluate the performance of \textit{MedMatch} via an extensive simulation study, and apply it to estimate a causal ERF for long-term \PM\ exposure and respiratory hospitalization in children using nationwide Medicaid data. 



\vspace{-.25in}

\section{Methods} \label{Section:Methods}

\subsection{Notation and estimand}

The notation below follows \cite{wu2022matching}. Let $j=1,...,N$ index the units of analysis (zip codes in our motivating data). Suppose $W_{j}$ denotes the observed continuous exposure for unit $j$, and $\mathbf{C}_{j}=(C_{1j},\ldots,C_{pj}) \in \mathbb{R}^p$ denotes a vector of $p$ measured covariates. Let $Z_{j}$ be a special continuous covariate that has to be measured in traditional GPS matching approaches, but in our proposed method, we will allow it to remain unmeasured while utilizing measured surrogates as substitutes. Two surrogate variables are measured - a continuous unit-level surrogate of $Z_j$, denoted by $Z^*_j$, and a continuous cluster-level variable $U_j$ (its value differs for units across but not within clusters) that induces heterogeneity in the distribution of $Z$. In our motivating Medicaid data, $Z_j$ is the unmeasured zip code-level MHI-MC, $Z^*_j$ is the measured MHI of the total zip code population, and $U_j$ is the Medicaid eligibility threshold of the state containing each of the zip codes, which is common across all zip codes within a state and a Medicaid eligibility cluster. 
$Y_{j}$ is the observed outcome, assumed here to be continuous. Under the standard Stable Unit Treatment Value Assumption (SUTVA) for causal inference, we define $Y_{j}(w)$ to be the counterfactual outcome for unit $j$ at exposure level $w$ \citep{rubin1974estimating,rubin1980randomization}. 
Let $f (w\mid\mathbf{c},z)$ be the true GPS, i.e., the conditional probability density function of exposure level $w$ given covariates $\mathbf{c}$ and $z$. 
We aim to estimate the population average causal ERF $\mu(w) = E\left[ Y_j(w) \right]$ for exposures $w \in \mathbb{W}$, where the expectation is taken over the distribution of counterfactual outcomes in the population of interest.

\subsection{Causal ERF identifying assumptions with surrogates}\label{Section:Assumptions}

We first present assumptions that are sufficient to allow the causal ERF to be identified using the surrogate variables. 
Recall the relationships between these variables posited in Section~\ref{Section:Introduction}: $Z^*$ is a rank-preserving function of $Z$ within levels of $U$. In this setting, the standard identifying assumptions for causal ERFs must be expanded slightly so that unconfoundedness and identifiability can be obtained when relying on $Z^*$ and $U$ rather than $Z$. In particular, we must make the conventional causal consistency, positivity, unconfoundedness and smoothness assumptions (conditional on the true $Z$) exactly following \cite{wu2022matching}, which enable identifiability of the ERF via conditioning on the GPS. These are given in Web Appendix A.1 and referred to as Assumptions 1-4. We also need the following two additional assumptions to enable identifiability on the GPS constructed using the surrogates.


\begin{assumption}[Relationship between $Z,Z^*,U$]
    $Z^*=g_u (Z)$ where $g_u$ is a deterministic, injective function for a given $U=u$.
\end{assumption}

\noindent Assumption 5 states that within levels of $U=u$, the surrogate $Z^*$ is a injective function (one-to-one) of the true $Z$. Any function that is rank-preserving is also injective, so that this assumption is satisfied if $Z^*$ is a rank-preserving function of $Z$ within levels of $U$.

\begin{assumption}[Balancing Condition of $U$]
    $W \, \indep \, U  \, | \, C,Z$
\end{assumption}

\noindent Assumption 6 states that $U$ is independent of $W$ after conditioning on the true confounders $C,Z$. Note that this assumption cannot be empirically verified since we have not measured $Z$, but intuitively  after conditioning on the true confounders, there should be no association between \PM\ and Medicaid eligibility thresholds. 

Under these assumptions, the density of $W$ conditional on $Z^*$ and $U$ is equivalent to the density conditional on the true confounder $Z$, i.e.,
\begin{align*}
    f(w|c,z)&=f(w|c,z,u) \quad \quad \text{by Assumption 6}\\
    & = f(w|c,z^*,u) \quad \quad \text{by Assumption 5}
\end{align*}

\noindent Note that the second line holds because, given some $U=u$, under Assumption 5 $Z=z \iff Z^*=z^*$ for some $z^*$. Since the density of $W$ conditional on $Z$ is equivalent to that conditional on $Z^*$ and $U$, identifiability under the GPS using $Z$ implies identifiability on the GPS using $Z^*$ and $U$. Thus, under these identifying assumptions, simply conditioning on $Z^*$ and $U$ in the GPS model instead of $Z$ would, in principle, avoid introducing any bias due to the unmeasured $Z$. However, in the next section we argue that estimation of causal ERFs via GPS matching is often complicated by practical data limitations that may be particularly pronounced in our application, and we propose a modified approach that may perform better than simply including $Z^*$ and $U$ in the GPS model in this context.

\subsection{Practical considerations for ERF estimation with clustered data}\label{sec:practical}

To better motivate the practical issues that often arise with ERF estimation in finite sample settings, we first explain how GPS matching is traditionally operationalized (while the same issues are likely to affect all causal ERF estimation methods, they are most easily demonstrated via matching due to its intuitive nature). In the binary exposure case, propensity score matching is simplified because each unit only has a single counterfactual outcome, namely its outcome under the treatment level not received. We can approximate that counterfactual outcome using the outcome observed for a matched unit (with similar confounder values) that received the opposite treatment. The complexity of matching surges in the continuous exposure case, where each unit theoretically has an infinite number of counterfactual outcomes. To make this problem tractable, $L$ equally sized exposure windows are constructed based on a user-specified caliper $\delta$. The caliper defines the diameter of the neighborhood set for any exposure level $w$, i.e., $[\min(w), \min(w)+\delta], [\min(w)+\delta, \min(w) + 2\delta], \ldots, [\min(w)+(L-1)\delta, \min(w) + L\delta]$ where $\min(w)=\min_{\{j \in \{1,\ldots,N\}\}} w_j$. We refer to the mid points of the exposure windows as ``pseudo-exposure'' values, $w^{(l)}$.
We then seek to approximate the counterfactual outcome for each unit at each pseudo-exposure level $w^{(l)}$ by finding a matched unit that experienced exposure within a window of size $\delta$ centered at $w^{(l)}$.

Because GPS are not known in most observational data settings, a GPS model is typically posed and a GPS is estimated for each unit at each pseudo-exposure level. For observed unit $j$, the matching function proposed by \cite{wu2022matching} finds a matched unit $j'$ for exposure level $w^{(l)}$ with $W_{j'}\in [w^{(l)}-\delta,w^{(l)}+\delta]$ and that minimizes a weighted average of distance between the estimated GPS of units $j$ and $j'$ and distance between $w^{(l)}$ and the observed exposure for unit $j'$, on a standardized scale. Intuitively, the aim is to select a matched unit that had similar probability of receiving exposure $w^{(l)}$ to that of unit $j$ (measured by the distance between the estimated GPS of the two units at/near pseudo-exposure level $w^{(l)}$) and that had an observed exposure level close to $w^{(l)}$. This matched unit's outcome is then used as an approximation of the outcome that unit $j$ would have experienced under exposure $w^{(l)}$, i.e., its counterfactual outcome at exposure $w^{(l)}$.
Using this procedure, and matching with replacement, we identify matches for all units in the data at each of the $L$ pseudo-exposure levels, yielding a matched dataset of size $N \times L$.
A smoothed average ERF is then obtained by applying a kernel smoother on the relationship between $W$ and $Y$ in the matched dataset \citep{wand1994kernel,kennedy2017non}.

One of the key practical obstacles to causal inference with continuous exposures is the high likelihood of data sparsity in some areas of the exposure range, even when sample sizes are very large. This sparsity often results in what is referred to as ``non-overlap''. Non-overlap occurs when, by chance, at some exposure level the sample contains few or no units representing certain areas of the confounder space. This is the empirical analog of a positivity violation, which occurs when units in some parts of the confounder space are ``ineligible'' to receive certain exposures. If not addressed, non-overlap generally leads to residual confounding in the matched data, i.e., the availability of few/no appropriate matches for some units at some exposure levels causes exposure-confounder imbalance in the matched data as well
\citep{nethery2019estimating}. Moreover, even if confounders are able to be balanced via matching, data sparsity in some areas of the exposure range can induce bias due to imbalance of precision variables and/or effect modifiers (not associated with $W$, so not confounders) in the matched data. This can occur because, due to the need to find matches for all units in the data from among a small pool of candidates at a given exposure level, a few units may end up serving as matches for a huge number of other units. These few units that are given very high weight at the given exposure level may, by chance, have a different distribution of some precision variables and/or effect modifiers than the sample. This leads to distortion of the distribution of these variables at some exposure levels in the matched data, effectively inducing a confounding effect of these non-confounders post-matching. In applications of GPS matching, truncation of the exposure range prior to ERF estimation has been used to reduce the influence of data-sparse areas \citep{wu2020evaluating,wu2022matching}. However, to our knowledge there have been no general recommendations for how to determine when/where to truncate. 

The likelihood of bias induced by the data sparsity/non-overlap problem may be exacerbated in the clustered data setting, which we illustrate using our motivating application. Strong geographic patterns in both \PM\ and Medicaid eligibility thresholds create significant differences in the distribution of \PM\ across clusters. In particular, the Medicaid eligibility threshold, $U$, is shared for all units in a given state (and cluster), and strong regional patterns exist in \PM\ such that some states rarely/never experience \PM\ concentrations in some areas of the exposure range. This is clear from Figure~\ref{Figure:StackedHist}, which shows the distribution of \PM\ stratified by cluster. This figure also confirms that even extensive truncation of the exposure range is unlikely to yield a scenario where all clusters are well-represented at all levels of \PM. The limited representation of certain clusters in certain parts of the exposure range means that distribution of $U$, and thereby of $Z$, in the sample is unlikely to be preserved at each exposure level in the GPS-matched data. As discussed above, imbalance in the distribution of these variables across exposure levels in the matched data will bias the estimated ERF. Thus, while simple inclusion of $Z^*$ and $U$ in the GPS model is theoretically sufficient to account for the confounding and/or effect modifying influence of the unmeasured $Z$, practical data limitations may compromise this capacity. To reduce the impact of this issue, in the next section we propose \textit{MedMatch}, which modifies the GPS matching target to not only seek matches with similar GPS values but also to explicitly encourage matching of units in the same cluster or across clusters with similar values of $U$. This is intended to yield matched data that better maintains the sample's distribution of $U$ and $Z$, particularly in data-sparse areas of the exposure range.



\vspace{-.15in}

\subsection{Proposed \textit{MedMatch} Method} \label{Section:Medmatch_Methods}

In this section, we describe our proposed alternative matching approach, called \textit{MedMatch}, for the estimation of a population average causal ERF in the presence of surrogates and cluster-patterned data sparsity. \textit{MedMatch} utilizes analogous procedures to the standard GPS matching algorithm but estimates the GPS using the surrogates, $Z^*$ and $U$, rather than the true but unmeasured confounder $Z$, and defines an alternative matching target function to encourage matching within clusters or across clusters with similar $U$ values.



\textit{MedMatch} first estimates the GPS using the measured confounders and the surrogates. For ease of notation, let $\tilde{\mathbf{c}}_j=\left( \mathbf{c}_j , z^*_j, u_j \right)$ be the vector of the observed confounders and surrogates for unit $j$ and $e_j (w,\tilde{\mathbf{c}}_j)=f(w\mid \tilde{\mathbf{c}}_j )$ denote the GPS at $W=w$ for unit $j$. Following \cite{wu2022matching}, we assume a Normal conditional density for the GPS. The conditional expectations $E(W_j\mid \tilde{\mathbf{C}}_j)$ are estimated using tree-ensemble models to allow for flexible non-linear associations between variables. The variance is assumed to be constant and is estimated via the sample variance of the tree ensemble residuals. The GPS estimates $\hat{e}_j (w,\tilde{\mathbf{c}}_j)$ are obtained by plugging in estimates of the conditional means and variance into a Normal density function.

To conduct matching, we first define exposure windows (using a caliper $\delta$) and pseudo-exposure levels $w^{(l)}$, as described in the previous section. For each $w^{(l)}$, the matching function selects a matched unit $j(w^{(l)})$ for each unit $j=1,\ldots,N$ based on two criteria: 1) to ensure the matched unit has observed exposure near $w^{(l)}$, its observed exposure $w_{j(w^{(l)})}$ must lie within $\delta$ caliper of $w^{(l)}$ (i.e., $w_{j(w^{(l)})} \in [w^{(l)}-\frac{1}{2}\delta, w^{(l)}+\frac{1}{2}\delta]$) and 2) to ensure two units are comparable in terms of both estimated GPS and $U$, the matched unit must be the nearest neighbor of the observed unit $j$ (among the subset of units meeting criterion 1) based on a two-dimensional distance calculated using the estimated GPS and $U$, on a standardized scale. Specifically, we define the \textit{MedMatch} matching function as follows: 
\begin{equation}\label{eq:MedMatch}
    j(w^{(l)}) =\, \text{arg} \ \underset{j':\mid w_{j'} - w^{(l)} \mid \le \delta}{\text{min}}\ \big\{  \tau \mid\mid (\hat{e}_{j}^*(w^{(l)},\tilde{\mathbf{c}}_j), \, \hat{e}_{j'}^*(w_{j'},\tilde{\mathbf{c}}_{j'}))\mid\mid + \, (1-\tau)\mid\mid(u^{*}_{j} , \, u^{*}_{j'})\mid\mid \big\}.
\end{equation}
where $\mid\mid \cdot \mid\mid$ is the distance metric (e.g. $L_1$ or $L_2$ distance), $\hat{e}^{*}_{j}(w^{(l)},\tilde{\mathbf{c}}_j)$ and $\hat{e}^{*}_{j'}(w_{j'},\tilde{\mathbf{c}}_{j'})$ are the normalized estimated GPS values (min-max normalization procedure described below) at the pseudo exposure $w^{(l)}$ for unit $j$ and at the observed exposure $w_{j'}$ for unit $j'$, respectively. $\tau \in [0,1]$ is a scale parameter (its selection is described in Section~\ref{Section:Hyperparam}) that weights between the estimated GPS and $U$, with very high values of $\tau$ favoring selection of matches with GPS very close to that of unit $j$, regardless of cluster, and low $\tau$ favoring matches within cluster. $\hat{e}^*_j$ and $u^*_j$ represent the min-max normalization versions of $\hat{e}_j$ and $u_j$ to put the two measures on a comparable scale , i.e., $u^{*}_j = \frac{u_j - \text{min}(u)}{\text{max}(u) - \text{min}(u)}$ and $\hat{e}^{*}_j = \frac{\hat{e}_j - \text{min}(\hat{e})}{\text{max}(\hat{e})- \text{min}(\hat{e})}$. We allow matching with replacement: a single unit can serve as a match for multiple other units. Then, the missing potential outcome of $j$ at pseudo-exposure level $w^{(l)}$ is imputed using the observed outcome of the matched unit $j(w^{(l)})$, i.e., $\hat{Y}_{j}(w^{(l)})=Y^{obs}_{j(w^{(l)})}$. 

After implementing matching, a smoothed average ERF $\hat{\mu}(w)=\hat{\mathbb{E}}[Y(w)]$ across the range of exposures $w\in [\min(w),\max(w)]$ is obtained by fitting a kernel smoother on the entire matched dataset where an optimal bandwidth was chosen to minimize the risk function described in \cite{kennedy2017non}. We use m-out-of-n bootstrap to construct a point-wise Wald confidence band for the causal ERF as proposed by \cite{wu2022matching}. Details of this procedure are provided in Web Appendix A.2.

\subsection{Diagnostics and Hyperparameter Selection}\label{Section:Hyperparam}

To assess possible residual confounding on measured variables, \cite{wu2022matching} recommend quantifying the linear relationship between exposure and each of the covariates in the matched data using the absolute correlation (AC) (detailed in Web Appendix A.2). If covariate balance is achieved via matching, the AC between the exposure and each covariate should be close to zero. We set the threshold for achieving covariate balance to be average of ACs less than $0.1$ \citep{zhu2015boosting}.

While achieving covariate balance and eliminating confounding in the matched data is critical, it is also important to ensure that the characteristics of the original sample are preserved in the matched data. This is particularly salient in our context due to the potential data sparsity issues discussed in Section~\ref{sec:practical}. Thus, to assess whether the distribution of measured variables (covariates, effect modifiers, and/or precision variables) in the original sample is well preserved after matching, we compute the Kolmogorov–Smirnov (KS) statistic, defined as the largest vertical difference between the empirical cumulative density functions of a given variable in the original data versus the matched data. The smaller the KS statistic, the closer together the distributions of the variable in the original and matched data.

Because both covariate balance and preserving distributions of key variables in the matched data are priorities, we propose using both AC and KS to select the hyperparameters. This deviates from the convention of using AC alone to select hyperparameters in GPS matching \citep{wu2022matching}. Specifically, we select hyperparameters $(\delta,\tau)$ that minimize the average of the normalized ACs and KSs across controlled covariates $\{\mathbf{C},Z^*,U\}$ using a grid search approach. Note that we first conduct min-max normalization of the ACs and KSs to put the two measures on a comparable scale (detailed in Web Appendix B.3).

\vspace{-.25in}

\section{Simulations} \label{Section:Simulations}

\subsection{Simulation settings}

The data generating mechanism for the simulation studies is informed by those in \cite{kennedy2017non} and \cite{wu2022matching}. Details on data generating processes are provided in Web Appendix B.1 and described briefly here. We generate $N=2,000$ samples falling into $G$ distinct clusters. We generate five measured covariates $\mathbf{C}=(C_1,\dots,C_5)$, an unmeasured true confounder variable $Z$, a surrogate $Z^*$ and a cluster-level surrogate $U$. We generate exposure, $W$, based on the GPS model with pre-exposure covariates $\mathbf{C}$ as well as the true $Z$. We generate outcome $Y$ from an outcome model specified as a cubic function of $W$ with confounding and/or effect modification by $Z$. 

We simulate data under four different cluster structure scenarios. In Scenarios 1 and 2, we generate $10$ and $50$ equally sized clusters (i.e. balanced clusters), each containing $200$ and $40$ units, respectively. In Scenarios 3 and 4, we generate $10$ and $50$ clusters, respectively, where the number of units in each cluster varies (i.e. imbalanced clusters). In Scenario 3, which most closely mirrors the real Medicaid data structure, the number of units in each cluster ranges from $14$ to $422$. In Scenario 4, the number of units in each cluster ranges from $13$ to $67$. Within each scenario, we obtain differing confounding and effect modification structures by varying $(\beta_{Z}, \beta_{WZ})$, where $\beta_{Z}$ represents the influence of $Z$ on the outcome $Y$, and $\beta_{WZ}$ represents the effect modifying influence of $Z$. $\beta_{Z}=0$ implies no confounding due to $Z$. $\beta_{WZ}=0$ implies no effect modification due to $Z$. 

\subsection{Competing methods and methods implementation}\label{Section:Extension_CausalGPS}

We apply \textit{MedMatch} to each simulated dataset as described in Section~\ref{Section:Medmatch_Methods}. To our knowledge, no other studies to date have attempted to estimate causal ERFs in a setting with clustered data and surrogate variables. However, here we formalize several simple adaptations of traditional GPS matching that might naturally be adopted in this setting, which will serve as competing methods in our simulation studies. 

One competing approach, which we refer to as \textit{adjusted}, applies conventional GPS matching with $\mathbf{C}$, $Z^*$, and $U$ included in the GPS model. Another competing approach, \textit{within}-cluster matching, conducts GPS estimation and matching entirely separately for each cluster (with $Z^*$ included in the GPS model), and the cluster-specific matched datasets are combined to estimate the population-level ERF. The \textit{adjusted} and \textit{within} approach should theoretically suffice for confounding adjustment, but may be subject to the practical limitations discussed in Section~\ref{sec:practical}. The \textit{fixed} approach emulates a scenario where the values of $U$ are not known exactly but the clustering structure is known, i.e., we know the sets of states that have the same $U$ value (in our motivating example, clusters are sets of states with the same Medicaid eligibility threshold). In the \textit{fixed} approach, the GPS model includes $\mathbf{C}$, $Z^*$ and cluster membership indicators (also known as ``fixed effects''). Each of the approaches uses a different GPS model specification but implements standard GPS matching 
\citep{wu2022matching} with hyperparameter selection as described in Section~\ref{Section:Hyperparam}. In each approach, we estimate the GPS conditional means using a fitted extreme gradient boosting machine (GBM) implemented in the \texttt{Superlearner} R package \citep{zhu2015boosting,van2007super}. More detail is provided in Web Appendix B.3. 

\subsection{Measures for assessment of the performance}

To assess the performance of the different methods, we calculated the absolute bias (AB) and root mean squared error (RMSE) of the estimated ERF \citep{kennedy2017non}. Moreover, we compare the KS statistic of all covariates, but with a particular emphasis on $Z$ and $U$, between the original data and the matched data to assess whether the sample's distribution of these variables is well preserved after matching. We further estimate the effective sample size (ESS) in the matched data, which corresponds to the ESS of a weighted sample. $ESS=1$ indicates that only one unit is used as a match and it serves as the match for every other unit in the data, whereas $ESS=N$ indicates that every unit is used as a match. Thus, a small ESS would indicate the possible existence of a few very influential observations with numerous matches \citep{chattopadhyay2020balancing}, which is generally undesirable as it leads to a high-variance estimator. Details on definition of AB, RMSE, the KS statistic, and ESS are provided in Web Appendix B.2.

\subsection{Simulation results}

\begin{figure}
     \centering
     \begin{subfigure}[b]{\textwidth}
         \centering
            \includegraphics[width=\textwidth,height=0.35\textheight,keepaspectratio]{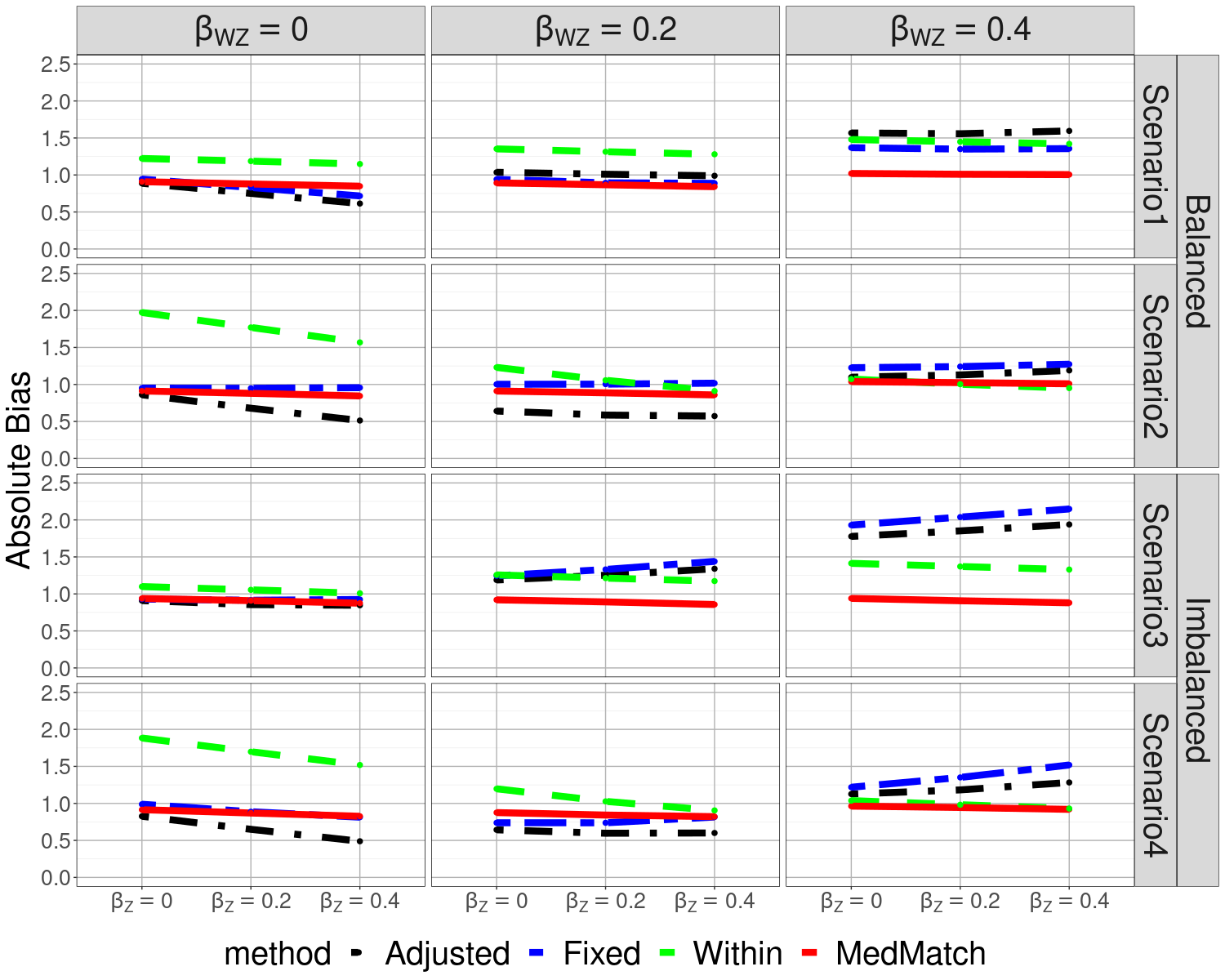}		
         \caption{Absolute Bias (AB)}
         \label{Figure:AC_KS_Plot_AB_MedMatch_corrUC_n2000_gpsspec1}
     \end{subfigure}\\
     \hfill
     \begin{subfigure}[b]{\textwidth}
         \centering
            \includegraphics[width=\textwidth,height=0.35\textheight,keepaspectratio]{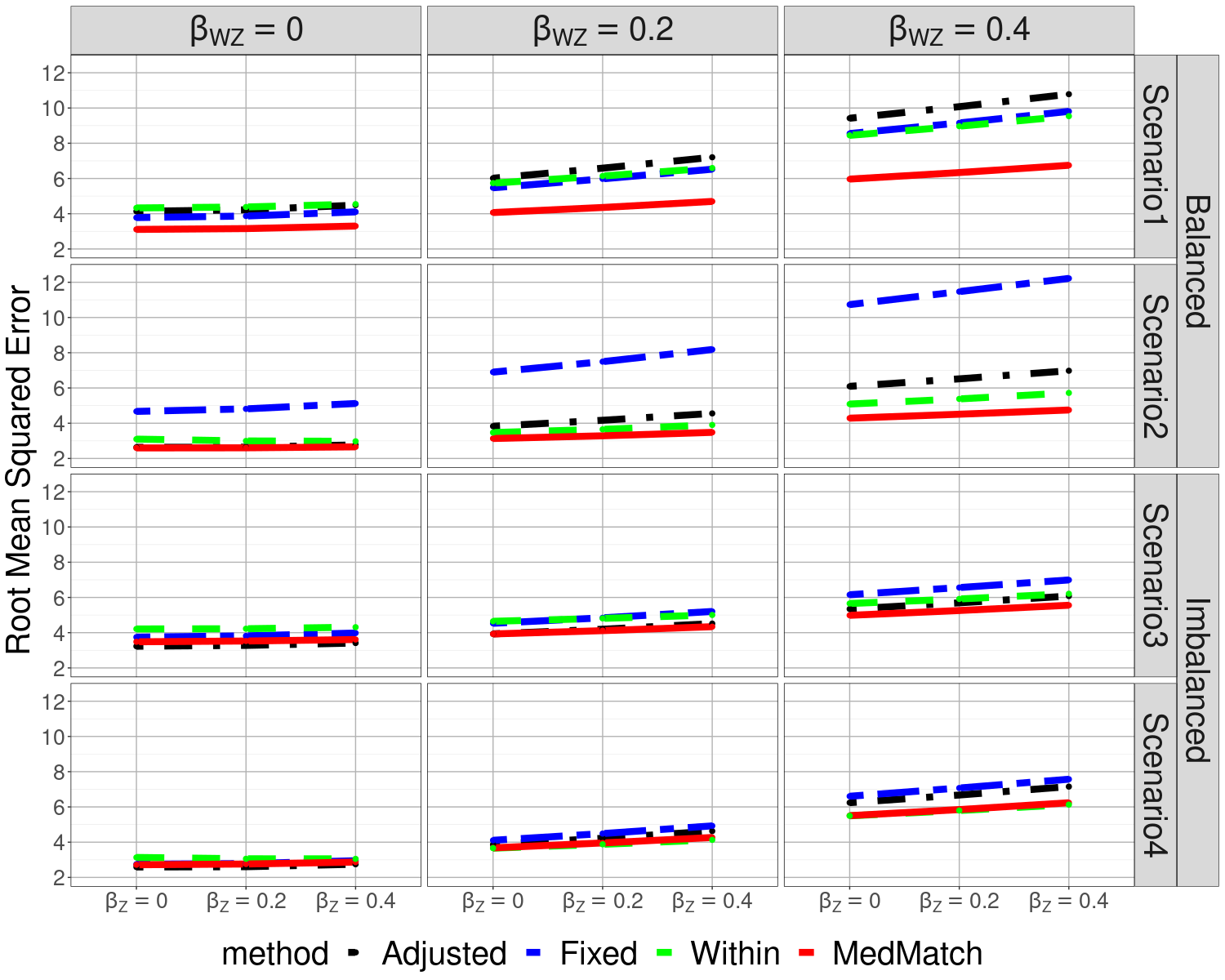}		
         \caption{Root Mean Squared Error (RMSE)}
         \label{Figure:AC_KS_Plot_MSE_MedMatch_corrUC_n2000_gpsspec1}
     \end{subfigure}
        \caption{Simulation results: absolute bias (a) and root mean square error (b) of the \textit{adjusted} (black dot dashed line), \textit{fixed} (blue dashed line), \textit{within} (green dotted line) and \textit{MedMatch} (red solid line) methods across 500 simulations. The simulated data are composed of $G=10$ equally sized clusters in Scenario 1; $G=50$ equally sized clusters in Scenario 2; $G=10$ clusters with varying cluster size in Scenario 3; and $G=50$ clusters with varying cluster size in Scenario 4. $\beta_{Z}$ and $\beta_{WZ}$ represent the strength of confounding and effect modification by $Z$, respectively.}
        \label{Figure:AC_KS_Plot_AB_MSE_MedMatch_corrUC_n2000_gpsspec1}
\end{figure}

\begin{figure}
     \centering
     \begin{subfigure}[b]{\textwidth}
         \centering
            \includegraphics[width=\textwidth,keepaspectratio]{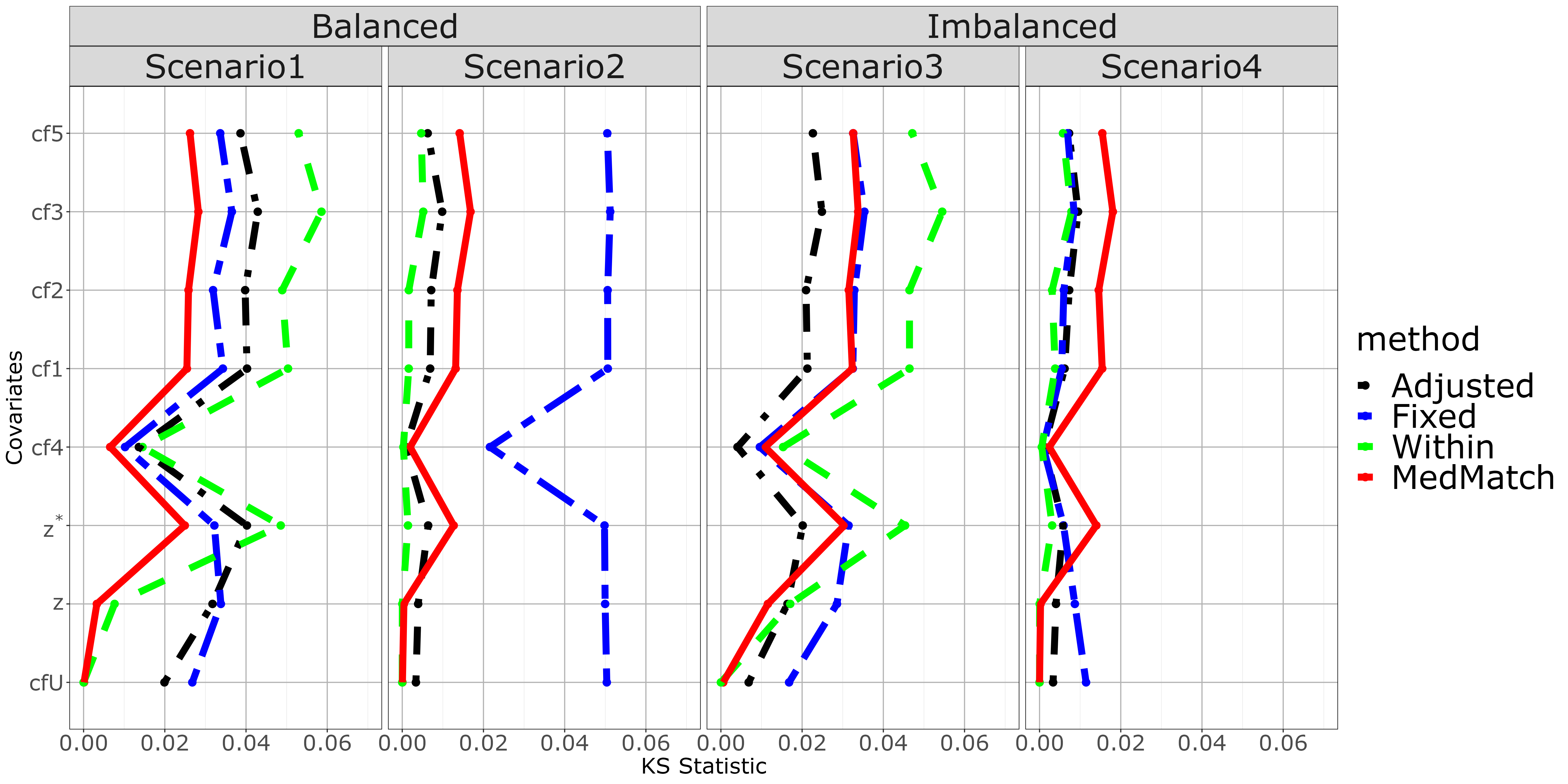}		
         \caption{KS distance}
         \label{Figure:KS}
     \end{subfigure}\\
     \hfill
     \begin{subfigure}[b]{\textwidth}
         \centering
            \includegraphics[width=\textwidth,keepaspectratio]{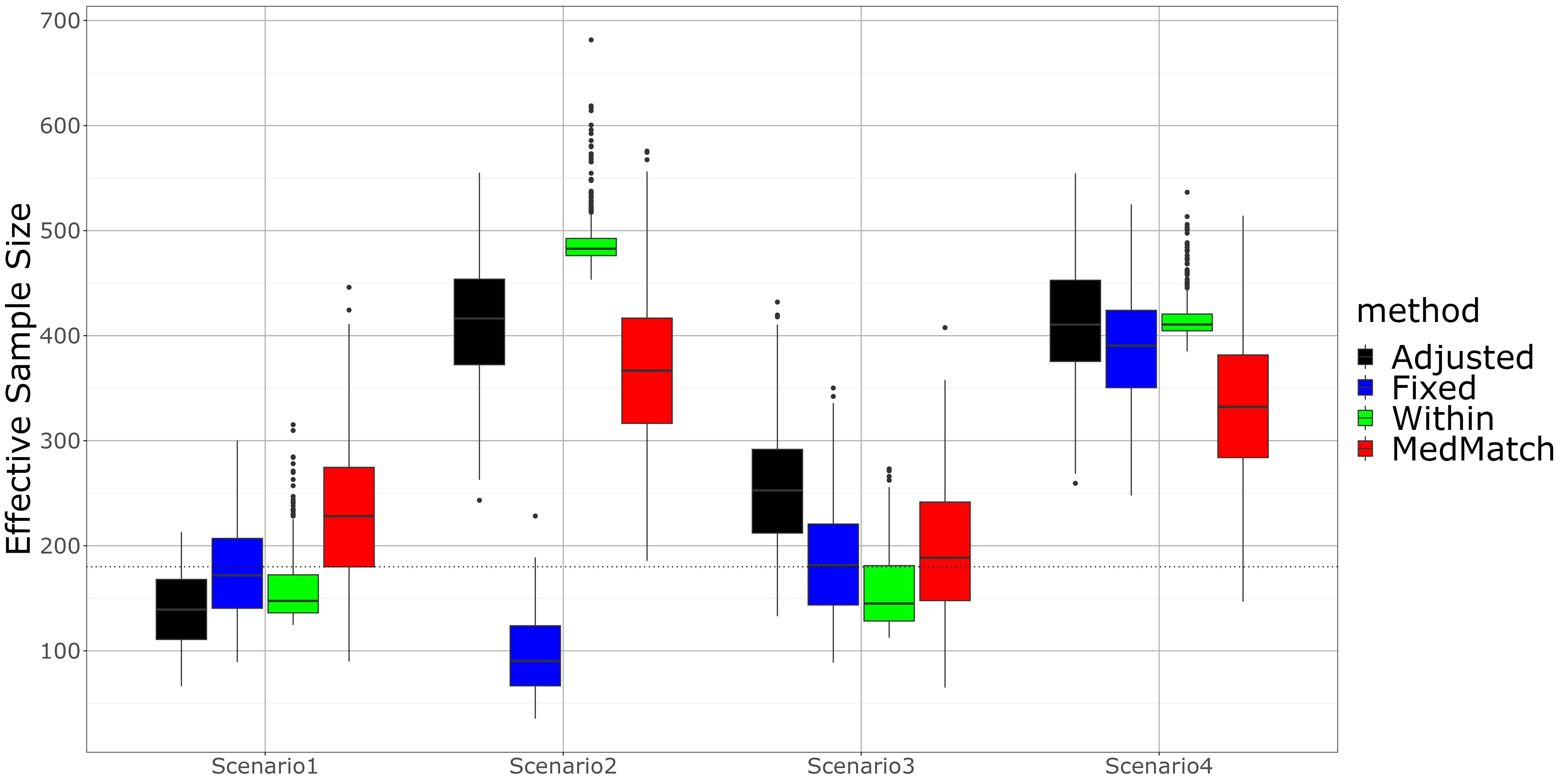}		
         \caption{ESS}
         \label{Figure:ESS}
     \end{subfigure}
        \caption{Simulation results: Average Kolmogorov-Smirnov (KS) statistic measuring distance between covariate distributions in the sample versus matched datasets (a) and distribution of effective sample sizes in the matched datasets (b) for the \textit{adjusted} (black), \textit{fixed} (blue), \textit{within} (green) and \textit{MedMatch} (red) methods across $500$ simulations in each of the four cluster structure scenarios.}
        \label{Figure:KSandESS}
\end{figure}

AB and RMSE results for each scenario are shown in Figure \ref{Figure:AC_KS_Plot_AB_MSE_MedMatch_corrUC_n2000_gpsspec1}. The average KS statistic for each covariate and the distribution of matched data ESS's across the simulations are plotted in Figure~\ref{Figure:KSandESS} (note that these measures are computed in the matched datasets but without using the outcome data, so that their values do not differ based on outcome model parameters $\beta_Z$ and $\beta_{WZ}$ as the AB and RMSE do). The optimal hyperparameters and the AC for each covariate can be found in Web Table B.1 and Web Figure B.4, respectively. Some consistent patterns can be observed across all three scenarios, which are described below.

First, we highlight general takeaways from the AB results. 
The strength of effect modification by $Z$ (size of $\beta_{WZ}$) impacts the relative performance of the methods more than the strength of confounding by $Z$ (size of $\beta_Z$). The AB of \textit{adjusted} and and \textit{fixed} tend to increase as the strength of effect modification increases, while the AB of \textit{MedMatch} remains small and nearly constant across all scenarios and confounding/effect modification structures (Figure~\ref{Figure:AC_KS_Plot_AB_MedMatch_corrUC_n2000_gpsspec1}).  
This is likely to be explained by the fact that the KS distance of $U$ (and of $Z$) in the original vs matched data is generally smaller when using \textit{MedMatch} compared to the other methods (Figure~\ref{Figure:KS}). This indicates that, as intended, \textit{MedMatch} better preserves the samples's distribution of $U$, and thereby of $Z$, in the matched data, despite $Z$ being treated as unmeasured.

Second, we highlight consistent patterns in the RMSE results (Figure~\ref{Figure:AC_KS_Plot_MSE_MedMatch_corrUC_n2000_gpsspec1}). \textit{MedMatch} always yields the smallest RMSE among all the methods in almost all scenarios. This is likely because \textit{MedMatch} has a consistently large ESS across the matched datasets (Figure~\ref{Figure:ESS}) and therefore consistently low variance, and it generally has the smallest bias among the methods as well. 

Comparing the results across scenarios provides insight into how cluster number, within-cluster sample size, and cluster balance impact the relative performance of the methods. We assess the impact of large or small cluster sample size in the balanced case (or imbalanced case) by comparing Scenarios 1 and 2 (or Scenario 3 and 4) (Figure~\ref{Figure:AC_KS_Plot_AB_MSE_MedMatch_corrUC_n2000_gpsspec1}). The performance of \textit{MedMatch} was consistently robust, while \textit{within} changes most notably across scenarios. For example, the AB of \textit{within} is largest when effect modification by $Z$ is not present ($\beta_{WZ}=0$) in small cluster sample size likely due to difficulty in balancing covariates $\mathbf{C}$ within each small cluster size and therefore residual confounding in the matched data (Figure 4 in Web Appendix B.5).  


Furthermore, we assess the impact of unbalanced cluster sizes by comparing Scenario 3 to 1 (or Scenario 4 to 2) (Figure~\ref{Figure:AC_KS_Plot_AB_MSE_MedMatch_corrUC_n2000_gpsspec1}). Relative to Scenario 1, in Scenario 3 the AB of \textit{fixed} generally increases, whereas the AB of the other methods remain relatively unaffected. This may be because the presence of some very small clusters leads to instability in the GPS model with fixed effects. On the other hand, the RMSE of all the methods are generally smaller in Scenario 3 than Scenario 1, due to generally higher ESS's in Scenario 3 (Figure~\ref{Figure:ESS}). Thus, on the whole, having balanced clusters does not seem to be advantageous for causal ERF estimation in clustered data. 

In addition, we assessed the impact of the choice of hyperparameter selection criterion on performance of each approach. Results of simulations comparing the methods under our proposed selection criterion combining AC and KS versus the convention of using only AC \citep{wu2022matching} are shown and discussed in detail in Web Appendix B.5. 

In summary, through simulation studies, we have shown that \textit{MedMatch} generally outperforms other possible extensions of existing methods in estimation of a population-level ERF in the clustered data with surrogate variables, especially in data structured to mimic features of the real Medicaid and \PM\ data. Moreover, we show its robustness to the number of clusters, cluster sizes, and varying strength of confounding and effect modifying influence of the unmeasured $Z$ variable.

\vspace{-.25in}

\section{\PM\ and Childhood Respiratory Outcomes in Medicaid} \label{Section:Medicaid}

Our study includes $23.4$ million children ages 6-18 enrolled in Medicaid during the period 2000-2012 and living in 46 states in the contiguous US and Washington, DC. Among the 48 contiguous US states, we exclude Kansas and Maine because Medicaid claims data for Maine from 2005-2010 and Kansas for 2010 are unavailable. Each beneficiary's residential zip code, state of Medicaid enrollment, age, race and sex are provided in the enrollment files. We use international classification of diseases (ICD) codes to identify inpatient hospitalizations where the primary or secondary cause was diseases of the respiratory system (ICD-9: 460-519) which includes pneumonia, influenza, and acute respiratory infection. We include only the first respiratory hospitalization for the enrollees who experienced multiple hospitalizations during the study period. We aggregate the individual-level data to the zip code level for each year to calculate respiratory hospitalization rates. Specifically, for each zip code and year, we divide the number of respiratory hospitalizations by the corresponding total person-years of follow-up in the zip code and use this rate as the outcome in our analyses.

We utilize the same zip code level annual \PM\ concentrations and potential confounders employed in several prior high-impact studies of the health effects of \PM\ \citep{di2017air, wu2020evaluating, wu2022matching, wei2022air}. In brief, we obtain daily predictions of \PM\ exposures on a 1-km$^2$ grid covering the continental US using an ensemble-based prediction model \citep{di2019ensemble}, which has been shown to have a cross-validated R$^2$ of 0.86. We average the daily \PM\ exposure predictions across grid cells within the boundary of each zip code and across the days within each year to obtain annual average \PM\ exposures, which are linked with the Medicaid data by zip code \citep{wu2020evaluating, wu2022matching, wei2022air}. We consider $18$ possible zip code level confounders such as socio-demographic, health behavior, and meteorological variables collated from various sources explained in Web Appendix C.1. 

Moreover, we obtain publicly available state-level Medicaid family income eligibility threshold as a percent of the federal poverty level (\%FPL) for age group 6-18 provided for each year \citep{brooks2019medicaid}. In summary, although our data include 46 states and Washington DC, some states have the same eligibility threshold, resulting in only 10 unique Medicaid eligibility thresholds (Figure~\ref{Figure:motivatingdata}) and therefore 10 clusters. 
Following previous implementations of GPS matching \citep{wu2020evaluating}, we truncate the exposure range to avoid the severe and possibly intractable data sparsity issues in the extremes of the exposure distribution. Specifically, we only include units (zip code-years) with observed exposures within the subset of the exposure range where at least 7 out of the 10 clusters are represented (see Web Figure C.1). This includes units with annual average \PM\ values within the range of $2.59 - 13.59$ \mugm. A detailed description of the data cleaning procedures can be found in Web Appendix C. Our final analytic data contain a total of $259,022$ zip code-years ($28,687$ unique zip codes).


To apply \textit{MedMatch} to the real data, we first estimate the GPS using a GBM and including all the confounders listed above as predictors as well as the state Medicaid eligibility threshold. 
We perform hyperparameter selection via a grid search over the hyperparameters using the combined AC and KS criterion. The selected hyperparameters were $(\delta,\tau)=(0.38,0.6)$. \textit{MedMatch} resulted in significant improvement in the covariate balance of the majority of covariates in the matched data relative to the original data (Web Figure C.2). The average AC improved from $0.14$ prior to matching to $0.06$ after matching. 

After matching, we estimate the causal ERF by fitting a kernel smoother with Gaussian kernel on the matched dataset. We use the block m-out-of-n boostrap procedure (Web Appendix A.2) to construct a 95\% confidence band. Specifically, for each re-sampling, we include all years associated with each sampled zip code to preserve across-year correlation. We sample $m=2\sqrt{n}\approx 338$ zip codes with replacement and we ensure that each cluster is adequately represented by sampling a fixed number of units from each cluster, proportional to its representation in the original dataset.

\begin{figure}
    \centerline{
    \includegraphics[width=\textwidth,height=0.4\textheight,keepaspectratio]{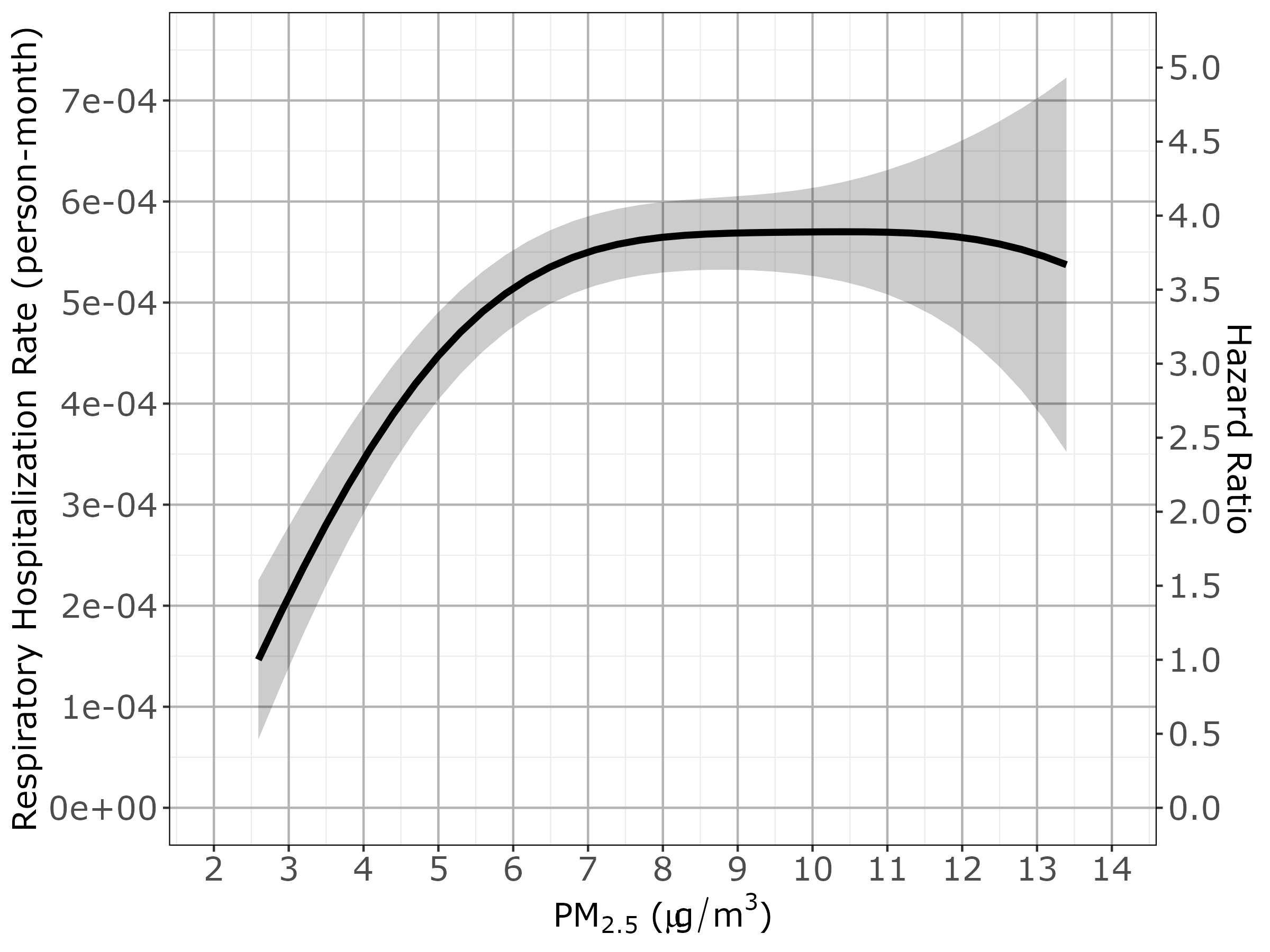}}
    \caption{The \textit{MedMatch} estimated causal ERF relating respiratory hospitalization to long-term \PM\ exposure in children enrolled in Medicaid (2000-2012). The shaded area represent the pointwise Wald 95\% confidence band, with standard errors calculated by block m-out-of-n bootstrap.}
    \label{Figure:Medicaid_ERF}
\end{figure}

Figure~\ref{Figure:Medicaid_ERF} shows the estimated average causal ERF relating long-term \PM\ exposure to respiratory hospitalization rate using \textit{MedMatch}. The right y-axis represents the associated hazard ratios which is calculated by dividing the outcome rate at each exposure level by the baseline rate (the estimated average outcome rate at the minimum \PM\ exposure value in the analytic data, 2.59 \mugm). We find a harmful effect of long-term \PM\ exposure on respiratory hospitalization rate in Medicaid children, as indicated by the increasing ERFs. Specifically, the curve is steeper at \PM $\le 8$ \mugm\ and starts to level off at higher concentrations. 
We also compare the causal ERF estimated using \textit{MedMatch} to \textit{adjusted} (Web Figure C.3). \textit{adjusted} and \textit{MedMatch} produce similarly steep curves at \PM $\le 8$ \mugm. However, the \textit{adjusted} curve continues to increase at higher concentrations, such that the estimated respiratory hospitalization rate at \PM=$13.39$ \mugm\ is $5.6$ times higher than the baseline rate, compared to $3.9$ times higher for \textit{MedMatch}. These differences may be driven by the fact that the KS distances are generally smaller when using \textit{MedMatch} compared to \textit{adjusted}, including for the Medicaid eligibility threshold (Web Figure C.2). This indicates that, as intended, \textit{MedMatch} better preserves the sample's distribution of potential confounders and effect modifiers. Moreover, \textit{MedMatch} gives smaller AC for each covariate in general and larger ESS than \textit{adjusted} (Web Figure C.2). 
 
\section{Discussion} \label{Section:Discussion}

In this paper, we proposed a causal inference method called \textit{MedMatch} to address challenges to estimating ERFs of environmental exposures on health outcomes using nationwide Medicaid claims data. \textit{MedMatch} accounts for the following unique features of nationwide children's Medicaid claims and environmental exposure data: 1) the data have a clustering structure due to differing income-based Medicaid eligibility thresholds by state, 2) MHI-MC, an important potential confounder and effect modifier, is unmeasured, and 3) two surrogate variables are measured, zip code MHI and state eligibility threshold. MHI was posited to be a rank-preserving function of MHI-MC, conditional on Medicaid eligibility threshold. Through simulation studies, we demonstrated that, under the requisite assumptions, \textit{MedMatch} is able to recover the population average causal ERF with little bias using the surrogate measures, and it consistently out-performs competing alternative approaches under a range of scenarios. The benefits of \textit{MedMatch} were most pronounced when the unmeasured variable is an effect modifier and a confounder, which is known to be the case for our income and air pollution motivating example. Moreover, we proposed a new criterion to find optimal hyperparameters and showed through simulation studies that it not only improves the performance of \textit{MedMatch} but also of existing GPS matching methods. 

We applied \textit{MedMatch} to estimate the causal ERF of long-term \PM\ exposure on first respiratory hospitalization in Medicaid children enrollees of age 6-18 between 2000 and 2012. Previous studies used nationwide Medicaid data and found positive associations between \PM\ exposure and health, similar to our finding. In particular, they reported positive associations between short-term \PM\ and cardiovascular and asthma hospitalizations in adults \citep{wei2022air, desouza2021nationwide} and between long-term \PM\ and asthma in children \citep{keet2018long}. Most other studies using Medicaid data have focused on single cities only and on short-term \PM\ exposure \citep{li2011association,wendt2014association, mann2010short}. To our knowledge, no previous studies have examined the ERF of long-term \PM\ exposure and respiratory hospitalization in children in Medicaid nationwide using a causal inference approach. Moreover, none of these papers has tackled the methodological challenges associated with the geographically-varying enrollment criteria for Medicaid, which generates a unique clustering structure in the data.

There are several strengths of this work. First, \textit{MedMatch} enables estimation of a population average causal ERF in the presence of an unmeasured confounder and/or effect modifier and cluster-patterned data sparsity, both of which occur frequently in real data settings. 
Second, compared to previously proposed matching methods for clustered data in the binary exposure setting \citep{arpino2016propensity,rickles2014two}, which require defining an arbitrary threshold to decide whether a candidate within-cluster match is ``appropriate", the proposed method instead enables selection of the weight $\tau$ for the proposed matching method through a data-driven approach. Third, the proposed matching approach can be straightforwardly extended to adjust for unmeasured spatial confounding by replacing $U$ in the matching function with a variable that measures spatial proximity, as \cite{papadogeorgou2019adjusting} proposed in the binary treatment setting.

There are also several limitations of our analyses that could be addressed in future work. First, \textit{MedMatch} assumes units in a cluster share a common eligibility criteria based on family income, which is generally the case only for children. To extend the analysis to the entire Medicaid population where different age groups have different eligibility criteria, further methods development may be needed. 
Second, although our assumption that $Z^*$ is a rank-preserving function of $Z$ is expected to be roughly met in our data application, it is likely not strictly true and is not testable. Third, \PM\ concentrations are measured only at centrally located monitors, necessitating the use of model-based \PM\ exposure predictions. These are subject to exposure measurement error; however, \cite{josey2023estimating} show that the measurement error in the predictions used here do not significantly impact causal ERF estimates. Fourth, because pollution sources at one location affect pollution and health at other locations, SUTVA may be violated in these data. Future research in the setting of exposure interference may be interesting.


\vspace{-.1in}

\section*{Acknowledgements}
This work was supported by NIH grants T32ES007142 and K01ES032458, the Harvard Data Science Initiative, and the Harvard Global Health Institute Burke Climate and Health Fellowship, in collaboration with the Salata Institute for Climate and Sustainability at Harvard University.

\section*{Data and code availability}
Software and R code to reproduce all analyses is available in \url{https://github.com/jennyjyounglee/MedMatch}. The Medicaid claims data used here can be obtained upon request and approval from the US Centers for Medicare and Medicaid Services. \PM\ data are publicly available at NASA’s Socioeconomic Data and Applications Center, and confounder data are publicly available from the US Census Bureau, Gridmet, and the US CDC. State-specific Medicaid eligibility threshold data are available on the Kaiser Family Foundation website at \url{https://www.kff.org/state-category/medicaid-chip/}. The computations in this paper were run on the Faculty of Arts and Sciences (FAS) Secure Research Computing Cannon cluster and FAS Secure Environment supported by the FAS Division of Science Research Computing Group at Harvard University. Analyses were conducted using R software version 4.0.5. 

\vspace{-.2in}

\bibliographystyle{apalike}

\end{document}